\documentclass[aps, prl, reprint]{revtex4-2}
\usepackage[hidelinks]{hyperref}

        \hypersetup{
           breaklinks=true,   % splits links across lines
           colorlinks=false,   % displays links as colored text instead of blocks
           pdfusetitle=true,  % \title and \author values into pdf metadata
        }

\usepackage[autostyle, english = american]{csquotes}
\MakeOuterQuote{"}
\usepackage{amsthm}
\usepackage{mathrsfs}
\usepackage{amssymb}
\usepackage{bm}
\usepackage{physics}
\usepackage{graphicx}

\usepackage{tensor}

\usepackage[scr = boondoxo, cal = stixplain]{mathalfa}
\usepackage[english]{babel}
\usepackage{amsmath}

\newcommand{\qgh}{ {\tilde {q_{k}}}}
\newcommand{\Gk}{:G_k:}
\newcommand{\qk}{q_{k}}
\newcommand{\g}{\bar g}
\newcommand{\R}{\bar R}

\newcommand{\h}{\hat h}
\newcommand{\cc}{\hat c}
\newcommand{\ac}{\hat{\bar c}}
\newcommand{\bb}{\hat b}

\begin{document}

\title{Asymptotic safety in Lorentzian quantum gravity}
\author{Edoardo D'Angelo}
\email{edoardo.dangelo@edu.unige.it}
%\date{\today}
\affiliation{Dipartimento di Matematica, 
Dipartimento di Eccellenza 2023-2027,
Università di Genova, Italy}
\affiliation{Istituto Nazionale di Fisica Nucleare -- Sezione di Genova, Italy}

\begin{abstract}
A recently introduced functional Renormalization Group (RG) provides a new tool to explore non-perturbative and covariant RG flows in Lorentzian spacetimes. We apply it for the first time to investigate the ultraviolet limit of quantum gravity. While the RG flow is state-dependent, it is possible to evaluate state and background independent contributions to the flow. Taking into account only these universal terms, the RG flow exhibits a non-trivial fixed point in the Einstein-Hilbert truncation, providing a mechanism for Asymptotic Safety in Lorentzian quantum gravity.
\end{abstract}

\maketitle

General Relativity (GR) was discovered in 1915; Quantum Mechanics in 1926. The realisation that the gravitational field should have been quantised along the same lines of the electromagnetic field came almost immediately: already in 1916, Einstein pointed out that quantum effects would modify the theory of General Relativity \cite{Einstein1916, *Stachel1999}.
The search for a consistent quantum theory of gravity has been fascinated generations of physicists ever since.

Among many conceptual puzzles, the main technical difficulty  in the quantization of gravity is that the standard approach of Quantum Field Theory (QFT) produces a quantum theory of gravity that is perturbatively non-renormalisable \cite{tHooft1974, GoroffSagnotti1985, GoroffSagnotti1985b}.

Perturbative non-renormalisability still leaves open the possibility of the Asymptotic Safety (AS) scenario \cite{Weinberg1976, Weinberg1980}, in which a QFT of the metric tensor is non-perturbatively renormalisable, thanks to the existence of a non-trivial fixed point in its RG flow. First realised in $2+\epsilon$ dimensions \cite{Christensen1978, Gastmans1978}, the AS scenario in four dimensions has been explored through lattice simulations \cite{Ambjorn2012, Loll2019}, and, in the continuum, through functional Renormalization Group (fRG) techniques \cite{Reuter1996, Berges2000, Niedermaier2006, ReuterSaueressig2012, Percacci2017, Bonanno2020, Saueressig2023}. While lattice computations are based on a background-independent regularisation of the Lorentzian path integral, fRG approaches are mostly based on the Euclidean formulation of the Wetterich equation \cite{Wetterich1992, Morris1993}, with few exceptions.

In 2011, an fRG-based approach to Lorentzian QG has been put forward, providing the first evidence of a non-trivial fixed point in the RG flow in Lorentzian spacetimes \cite{Manrique2011}. The computation was carried out assuming an ADM foliation of the background geometry and a compact time direction, which allowed for a resummation of Matsubara frequencies in the propagator. The Lorentzian fRG based on the ADM formalism initiated a study of AS in foliated spacetimes \cite{Biemans2017, SaueressigWang2023}. More recently, fRG-based investigations has been carried out for the graviton spectral function in Minkowski \cite{Fehre2021, Bonanno2021}. However, all fRG-based approaches in Lorentzian spacetimes had to give up background independence in favour of Lorentzian signature.

In this Letter, we provide the first evidence for a background-independent, non-trivial fixed point for quantum gravity in Lorentzian signature, in the Einstein-Hilbert truncation. The result is based on a novel Wetterich-type fRG equation (FRGE), directly developed in Lorentzian spacetimes with a covariant formalism and for any Hadamard state \cite{DDPR2022, DR2023}. This new RG equation uses a local regulator in position, thus acting as an artificial mass, and a Hadamard regularisation to subtract the UV divergences. Since it is written in terms of the interacting Feynman propagator, the Lorentzian FRGE exhibits state dependence \cite{DDPR2022}. The state is chosen for the free theory, and it acts as a background, fiducial quantity for the flow, similarly to the background geometry.

While a state for the graviton in general spacetimes is not known, here we show that the universal terms that must contribute in the FRGE for any state, and in all backgrounds, already determine the existence of a Reuter-type fixed point for Lorentzian quantum gravity.

\subsection{Quantum Gravity as a locally covariant QFT}

In order to apply the Lorentzian FRGE to gravity, we take as theoretical framework QG as a locally covariant QFT \cite{Brunetti2016, Brunetti2022b}. In this context, gravity is quantised on a fixed, globally hyperbolic spacetime $(\mathcal M, \bar g)$ with background metric $\bar g$; our computation is background independent in the sense that $\mathcal M$ is fixed, but arbitrary, thus studying the RG equations in all spacetimes at once \cite{Brunetti2001}.

The space of off-shell configurations is $\mathscr E(\mathcal M) = \Gamma( T^*(\mathcal M)^{\otimes 2}) \ni \h$, the space of symmetric bi-tensors. As usual, the configuration space must be extended to include the ghosts $\cc$, the antighosts $\ac$, and the Nakanishi-Lautrup fields $\bb $.
We collect an element of the extended configuration space in the field multiplet $\varphi := \{ \h , \cc , \bb , \ac \} \in \overline{\mathscr E}(\mathcal M)$.

In the Batalin-Vilkovisky (BV) formalism \cite{Batalin1977, Batalin1981, Batalin1983}, the configuration space is doubled to include the antifields, identified with the basis of the tangent space, $\varphi^\ddagger := \frac{\delta}{\delta \varphi}$. 
The classical BV algebra $\mathscr A$ is thus the algebra of local functions on the odd cotangent bundle of the extended configuration space \cite{FR, Fredenhagen2013}. 
The \textit{antibracket} is defined by $\{ \varphi_A(x) , \varphi^\ddagger_B(y) \} = \delta_{AB} \delta(x-y)$, with $A, B$ indices on the field space, and extended to functions of the fields and antifields by the graded Leibniz rule.

The dynamics is governed by the Euler-Lagrange equations of the action
\begin{equation}
I := I_{EH} + I_{af} + \gamma \Psi = I_{EH} + I_{af} + I_{gh} + I_{gf} \ ,
\end{equation}
where $I_{EH}= 2 \zeta^2 \int_\mathcal M \sqrt{-\det \hat g } ( R(\hat g) - 2\Lambda)$ is the Einstein-Hilbert action in terms of the full metric $\hat g := \bar g + \h$, and $\zeta^2 = (32 \pi G)^{-1}$ where $G$ is Newton's constant. The antifield term is
\[
I_{af} := \int_{\mathcal M} \sqrt{- \det \hat g } \mathcal {L}_{\cc} \hat{g}^{ab} h_{ab}^\ddagger + c^b \partial_b c^a \ c^\ddagger_a + i \bb^a \bar c^\ddagger_b \ ,
\]
where $\mathcal {L}_{\cc}$ is the Lie derivative. The gauge-fixing Fermion $\Psi$ in the De-Donder gauge is
\[
\Psi = i \int_\mathcal M  \sqrt{- \det \bar g} \ \ac^b (\nabla^a \h_{ab} - \frac{1}{2} \nabla_b \h_{ac} \bar g^{ac}) \ .
\]
Finally, the \textit{BRST differential} is defined as $\gamma := \{ \cdot , I_{af} \}$ \cite{BRS1974a, BRS1974b, BRS1975}, and the action satisfies the \textit{Classical Master Equation} $\{I , I \} = s I = 0$, where the \textit{BV differential} is $ s:= \{ \cdot , I \}$ \cite{FR, Fredenhagen2013, Brunetti2016, Brunetti2022b}.

Deformation quantisation proceeds splitting the action $I$ into a term quadratic in the fields $I_0$ and a remaining, interacting term $V:= I - I_0$. The free part $I_0$ is used to define the quantum products and the time-ordered products;
%in the classical BV algebra; 
the \textit{Epstein-Glaser renormalisation procedure} constructs the time-ordered products of local functions at coincidence points \cite{Brunetti2009, Brunetti1999, HollandsWald2001a, HollandsWald2001b, HollandsWald2004}.
% perturbatively, but without assuming power-counting: it thus works for gravity as well
Interacting observables are thus represented as formal power series in the $*-$algebra of free observables $\mathscr A$, and a state is a linear, positive, normalised functional $\omega: \mathscr A \to \mathbb C$ mapping the observable to its expectation value \cite{AAQFT15, Rejzner2016}.

In order to define the generating functionals, we introduce sources that couple linearly to the fields, $J:= \int_\mathcal M j_A \varphi^A$, and classical BRST sources that couple to their BRST variations $\Sigma:= \int_\mathcal M \sigma_A \gamma \varphi^A$. The contribution $\Sigma$ can be understood as a shift of the antifield term $I_{af}$, so that, even if evaluating on a state $\omega$ sets the antifield to the zero configuration $\varphi^\ddagger = 0$, the generating functionals still depend non-trivially on $\sigma$.

Finally, we need to introduce the regulator terms $Q_k$. These are chosen as local terms quadratic in the fields, acting as artificial masses in the correlation functions \cite{DDPR2022, DR2023}:
\begin{multline}\label{eq:regulator}
Q_k := - \frac{1}{2} \int_\mathcal M \sqrt{- \det \bar g(x)} \left [ T(\h_{ab}(x) \tensor{{\qk}}{^{abcd}}(x) \h_{cd}(x)) \right. \\ \left.
+ 2 T \ac_{a}(x)  \tensor{{\qgh}}{^{ab}}(x) \cc_b(x) \right ] \ ,
\end{multline}
where $T$ is the time-ordering operator. Notice that the regulator are local in position, preserving causality and Lorentz invariance. The regulator kernels $\qk$ and $\qgh$ are chosen proportionally to the RG scale $k$, and include a compactly supported smooth function $f(x) \in C^\infty_c$ which equals $1$ on a given region $\mathcal O \subset \mathcal M$.
%This in turn guarantees that the $S-$matrix is unitary.

Together with the regulator term, we also introduce a source for its BRST variation,
\begin{multline}
H(\eta) := \frac{1}{2}\int_\mathcal M \sqrt{- \det \bar g(x)} \left [ \eta(x) \ \gamma( \h_{ab}(x) \h^{ab}(x) ) \right.
\\ 
\left.
+ \tilde \eta \gamma( \ac_{a}(x) \cc^{a}(x) ) \right ] \ .
\end{multline}
Introducing a scale-dependent BV differential $s_k :=  s + \int_\mathcal M {\qk}_A \frac{\delta}{\delta \eta_A}$,
%, where the \textit{BV differential} is $s:= \{ \cdot , I \}$, 
the extended action $I_{ext} := I + \Sigma + Q_k + H$, satisfies a symmetry identity, extending the BV invariance of the classical action $I$ to \cite{DR2023}
\begin{equation} \label{eq:extended-CME}
s_k I_{ext} = 0 \ .
\end{equation}

The regularised generating functional for time-ordered correlation functions is defined as
\begin{equation}
Z_k(\bar g; j, \sigma, \eta) := 
%\omega \left( T(e^{-i T^{-1} V}) T(e^{i T^{-1}(V + \Sigma + J + Q_k + H) } \right ) = 
\langle T \exp{\Sigma + J + Q_k + H } \rangle \ ,
\end{equation}
in terms of the mean value $\langle F \rangle = \omega \left( T(e^{i V})^{-1} T(e^{i V} F ) \right )$. 
%Notice that the dependence on $\varphi^\ddagger$ and on $\sigma$ is not the same.

This definition of $Z_k$ generalises the usual path integral representation \cite{Weinberg-vol2}, to globally hyperbolic spacetimes and generic states $\omega$. In flat space, both in Lorentzian or in Euclidean signature, there is a unique Poincaré (or Euclidean) invariant ground state, and it is usually chosen to evaluate correlation functions. However, in curved spacetimes there is no unique choice for a vacuum (as it is known already for the scalar case, for example in Schwarzschild spacetimes \cite{Candelas1980}), and the choice of a state $\omega$ needs to be taken explicitly into account.

%A typical choice of state $\omega$ is the quasi-free state for the free theory which evaluates the field $\varphi$ in a particular field configuration; in this case, $\omega(F) = TF(0)$. The path integral typically refers to a heuristic representation of this state. Notice that, although the state is quasi-free for the free theory, $\omega(\varphi)$, this does not imply that the state is quasi-free for the interacting theory, since $\langle \varphi \rangle \neq 0$.

The Effective Average Action (EAA) $\Gamma_k (\bar g; \phi, \sigma, \eta)$ is defined in the standard way. The regularised generating functional for connected Green's functions $W_k(j):= \log Z_k$ defines the classical fields $\phi = W_k^{(1)} = \langle \varphi \rangle$ as functions of the sources $j$. The relation between the sources and the fields can be inverted in $W^{(1)}(j_\phi) = \phi$ \cite{DDPR2022}, so that the Legendre transform $\tilde \Gamma_k = W_k(j_\phi) - J_\phi(\phi)$ is well-defined. The EAA is the modified Legendre transform of the regularised generating functional of connected Green's functions $ \Gamma_k := \tilde \Gamma_k - Q_k(\phi)$. The EAA is thus a function of the classical fields $\phi := \{ h , c , b, \bar c \}$, of $\sigma$, and the scale $k$.

Thanks to the extended symmetry of Eq. \eqref{eq:extended-CME}, the Legendre EAA  satisfies the extended Slavnov-Taylor identity \cite{DR2023}
\begin{equation}
\int_\mathcal M \frac{1}{\sqrt{- \det \bar g(x)}} \left [ \frac{\delta \tilde \Gamma_k}{\delta \sigma_A(x)} \frac{\delta \tilde \Gamma_k}{\delta \phi^A(x)} + q_k^A(x) \frac{\delta \tilde \Gamma_k}{\delta \eta^A(x)} \right ] = 0 \ .
\end{equation}
The EAA is then constrained by the cohomology of the BRST operator $\gamma$ in ghost number zero; from the solution of the Wess-Zumino consistency condition \cite{Morris2022, Barnich1994-WZ}, it follows that the EAA must be a BRST-invariant functional of the full classical metric $g := \bar g + h$ \cite{DR2023}. 
%Moreover, The BRST sources $\sigma$ act as "mean antifields", in the sense that, in the space of mean fields $\phi$ and BRST sources $\sigma$, it is possible to define an antibracket and a BV differential, analogous to the BV formalism in the space of fields $\varphi$ and antifields. 

\section{Renormalization Group flow equations}
The RG flow equations for gravity are derived in complete analogy with the gauge theory case \cite{Reuter1993, DR2023}. 
They read
%The derivation starts from directly computing the expression
%\[
%\partial_k W_k = \langle \partial_k Q_k \rangle \ ,
%\]
%from which it immediately follows that
%\[
%\partial_k \Gamma_k = \langle \partial_k Q_k \rangle - \partial_k Q_k(\phi) \ .
%\]
%Expanding the definition of $Q_k$, it is immediate to see that
%\[
%\partial_k \Gamma_k = - \frac{1}{2} \int_\mathcal M \Tr{ \partial_k q_k \langle \lim_{x \to y} \varphi(x) \varphi(y) - h(x,y) \rangle_c } \ ,
%\]
%where $\langle \ \cdot \ \rangle_c$ denotes the connected part of the mean value operator.
%The subtraction of the Hadamard parametrix $h(x,y)$, encoding the divergences of the Feynman propagator in the coincidence limit, follows from the insertion of the time-ordering operator in the definition of the regulator, Eq. \eqref{eq:regulator} \cite{DDPR2022}. Thus, it is possible to define new counterterms $H_k$ by the commutation of the mean value operator with the limit,
%\[
%\langle \lim_{x \to y} ( T\varphi(x) \varphi(y) - h(x,y) ) \rangle = \lim_{y \to x} \langle T \varphi(x) \varphi(y) \rangle - H_k \ .
%\]
%The mean value can be re-expressed in terms of generating functionals, as $W_k^{(2)}(x,y) = \langle \varphi(x) \cdot_T \varphi(y) \rangle - \phi(x) \phi(y)$. Finally, recalling the standard relation between Legendre transforms
%\[
%(\Gamma_k^{(2)} - q_k) W_k^{(2)} = - 1 \ ,
%\]
%the FRGE reads
\begin{equation} \label{eq:RG}
\partial_k \Gamma_k(\bar g; \phi) = \frac{i}{2}  \int_\mathcal M \Tr{\partial_k \qk(x) \Gk(x,x) } \ .
\end{equation}
The trace is over Lorentz and field indices. The equations are written in terms of $\Gamma_k(\bar g, \phi) = \Gamma_k (\bar g, \phi, \sigma =0, \eta = 0)$, with the field $b$ integrated out, and the \textit{interacting propagator}, satisfying
\begin{equation}\label{eq:def-G_k}
 \frac{\delta^2}{\delta \phi(x) \delta \phi(z)}(\Gamma_k + Q_k) G_k(z,y) = - \delta(x,y) \mathbb I \ ,
\end{equation}
where $\mathbb I$ denotes an appropriate tensor identity.

Notice that, contrary to standard practice in the Asymptotic Safety literature in Euclidean space \cite{ReuterSaueressig2012}, the regulator terms are local in position. In Euclidean signature, the regulator is usually chosen to be a non-local function in position. This guarantees that the RG equation remains finite, without additional regularisations. However, in Lorentzian spacetimes there is no known example of a regulator that satisfies at the same time the requirements of Lorentz invariance, causality, and finiteness of the FRGE \cite{Braun2022}. In the case of cosmological backgrounds, an alternative is the use of a regulator depending on the spatial momenta only, since the background already breaks Lorentz invariance \cite{Banerjee2022}. 

Here, as the background is kept arbitrary, we choose a simple regulator local in position, $q_k(x) = k^2 f(x)$. Recall that $f \in C^\infty_c$, and equals $1$ on a region $\mathcal O \subset \mathcal M$. The advantage of such a regulator is that it preserves Lorentz invariance and causality. Moreover, the cut-off function $f$ acts as infra-red cut-off, since the r.h.s. of the FRGE \eqref{eq:RG} is proportional to $f$ itself. 

Ultraviolet finiteness is instead guaranteed by the normal-ordering prescription, arising from the definition of the EAA in terms of time-ordered quantities. It follows that the FRGE is both ultraviolet and infra-red finite by definition.

In fact, by direct computation one can see that the normal.ordered interacting propagator is given by 
\begin{align*}
\lim_{y \to x} :G_k: &= \langle \lim_{y \to x} ( T\varphi(x) \varphi(y) - h(x,y) ) \rangle \\
&= \lim_{y \to x} \langle T \varphi(x) \varphi(y) \rangle - H_k \ .
\end{align*}
The first expression arises from the insertion of the time-ordering operator in the regulator, Eq. \eqref{eq:regulator}. The commutation of the limit with the mean value operator thus defines the counterterm $H_k$, and it guarantees that FRGE are ultra-violet finite.

%While the counterterm $H_k$ arises from the definition of the normal-ordered quantity $\langle T\varphi(x)\varphi(y) \rangle$, it can be computed from the EAA itself. This is actually a crucial step, as it allows to close the FRGE as a functional differential equation for the EAA.

Operationally, the normal ordering of $:G_k:$ can be computed by a point-splitting procedure. 
Formally divergent quantities, such as $\langle \hat{h}^{ab}(x) \hat{h}_{cd}(x) \rangle = - i \tensor{{G_k^{hh}}}{_{ab}^{cd}}(x,x)$, are replaced by point-split expressions $\tensor{{G_k^{hh}}}{_{ab}^{a'b'}}(x,y)$, for $y$ in the vicinity of $x$ and space-like separated. The singular terms in the coincidence limit $H_k$ are subtracted, obtaining the regularised corresponding quantity $:\tensor{{G_k^{hh}}}{_{ab}^{ab}}: := G_k - H_k$.

Despite the use of a local regulator, it is still possible to prove that \cite{DDPR2022}
\[
\lim_{k \to \infty} \Gamma_k = I(\phi) + C \ ,
\]
where $C$ is a (finite) arbitrary constant. It follows that the EAA interpolates between the quantum action $\Gamma_0$ in the IR and the bare classical action $I$ in the UV. The FRGE thus describes an RG flow, even if strictly speaking it is derived as a flow of the EAA under rescalings of the mass parameter.

Mass-type regulators appeared already in the literature with the name of \textit{Callan-Symanzik cut-offs} \cite{Alexandre2001, Fehre2021}. The FRGE \eqref{eq:RG}, first derived in Refs. \cite{DDPR2022,DR2023}, shares some similarities with the recently introduced \textit{renormalised spectral flows} \cite{Braun2022}. The difference is in the definition of the counterterms $H_k$: in the case of renormalised spectral flows, they arise from the dependence of the regulator function $Q_k$ on an additional, UV cut-off scale $\Lambda$.

\subsection{State dependence}
In Lorentzian spacetimes, Eq. \eqref{eq:def-G_k} admits an infinite family of solutions $G_k$. This is the main difference from the Euclidean case, where the EAA admits a unique inverse. It follows that the FRGE depends on the choice of the interacting propagator $G_k$. Different propagators $G_k, G_k'$ differ by a smooth contribution $w-w'$, and they give raise to different RG flows. 

The ambiguity in the choice of the interacting propagator can be fixed choosing a state for the free theory $\omega$. Here, we recall the main argument and results; a detailed discussion can be found in Refs. \cite{DDPR2022,DP2023,DR2023}. Consider a region of the spacetime where the interaction is turned off, $V=0$. Then, the EAA reduces to $\Gamma_k(\phi) = Z_k I_0(\phi) + C$, for some finite constant $C$, where $I_0$ is the quadratic part of the bare action $I$ and $Z_k$ is a wavefunction renormalization. By direct computation the interacting propagator is then proportional to $\Delta_{F,k}$, the Feynman propagator for the free theory with a mass modified by the regulator term $q_k$. The Feynman propagator is fixed by the choice of a state $\omega$, as it is given by the time-ordered, connected two-point function. Moreover, if the state satisfies the Hadamard condition, the Feynman propagator has a universal UV singular structure $h_k$ \cite{AAQFT15, Radzikowski1996,Radzikowski1996b,KayWald}. It follows that the choice of a Hadamard state fixes the smooth contributions to the Feynman propagator $w_k := \Delta_{F,k} - h_k$.

When the interaction $V$ is turned on, the EAA can be decomposed into $\Gamma_k = I_0 + U_k(\phi)$, where $U_k$ incorporates all the quantum corrections, including non-local and higher derivative terms. The construction of the full interacting propagator $G_k$ follows from the free case by a perturbative-type construction, and in particular it is possible to prove that \cite{DDPR2022}
\begin{equation}\label{eq:G_k-series}
:G_k: = \sum_{n=0}^\infty (i \Delta_{F,k} U_k^{(2)})^n w_k \ .
\end{equation}
The series is uniquely fixed by the starting element $w_k$, and the requirement that $G_k$ is a fundamental solution for $\Gamma_k^{(2)} - q_k$. The interacting propagator and, by extension, the FRGE thus depend on the choice of the smooth contribution $w_k$, which uniquely fixes a quasi-free Hadamard state for the free theory, as quasi-free states are determined by their two-point function. In this way the FRGE inherits a dependence on the state for the free theory.

%This ambiguity reflects the fact that the EAA depends on the choi ce of a state: different states give raise to different correlation functions and thus to different physical behaviours. 

\section{Hadamard subtraction and Local Potential Approximation}

We now assume that the operator $\Gamma_k^{(2)} - q_k$ is Green hyperbolic, with the kinetic term, apart from a possible wavefunction renormalisation $Z_k$, given by the free part of the action: $\Gamma_k^{(2)} - q_k = Z_k D - q_k + U_k^{(2)}$, where $D= I_0^{(2)}$. In this approximation, the \textit{effective potential} $U_k^{(2)}$ does not contain derivatives of the Dirac delta.

In this case, it is known that the interacting propagator coincides with the propagator of the free theory, with a mass modified by $U_k$  \cite{Drago2015, DDPR2022}, by an exact resummation of the series in Eq. \eqref{eq:G_k-series}. This in particular means that, if $\Delta_{F,k}$ satisfies the Hadamard condition, $G_k$ is Hadamard as well. Thus, for $y$ in a normal convex neighbourhood of a given $x$, the interacting propagator must have the same Hadamard singularity structure of the free propagator:
\begin{equation} \label{eq:G_k}
G_k = \frac{i}{ 8 \pi^2\zeta^2_k} \left ( H_k  + W \right ) \ ,
\end{equation}
written in terms of a smooth contribution $W$ and the \textit{Hadamard parametrix}, capturing its universal UV singularity structure:
\[
H_k(x,y) =  \frac{i}{8 \pi^2 \zeta^2_k} \lim_{\epsilon \to 0^+} \bigg [ \frac{\Delta^{1/2}}{\sigma_\epsilon(x,y)} \mathbb I + V \log{\frac{\sigma_\epsilon(x,y)}{\mu}} \bigg ] \ .
\]
In the last equations, $\zeta^2_k := Z_k \zeta^2$, $\sigma(x,y)$ is the squared geodesic distance taken with sign between $x$ and $y$ and $\sigma_\epsilon(x,y) = \sigma(x,y) + i \epsilon$, $\Delta$ is the van-Vleck-Morette determinant. $\mathbb I$ is an appropriate tensor structure, depending whether $G_k$ describes the gravitational or the ghost propagator.

The distributions $V, \ W$ can be expanded in an covariant Taylor expansion as $V = \sum_{n=0} V_n \sigma^n$ and $W = \sum_{n=0} W_n \sigma^n$; the \textit{Hadamard recursion relations} determine higher orders in the expansion from the zeroth order \cite{DeWitt1960}. The zeroth term $V_0$ is completely determined by the quantum wave operator and the background geometry by the formula \cite{Hack2010}
\begin{equation}\label{eq:V0}
V_0 = - \frac{1}{2}\frac{\delta^2}{\delta \phi \delta \phi}(\Gamma_k + Q_k) \Delta^{1/2} \mathbb I \ ,
\end{equation} 
and the coincidence values $\Delta^{1/2}(x,x) = 1$, $\nabla_a \nabla_b \Delta^{1/2}(x,x) = 1/6 \R_{ab}$ \cite{Christensen1976}.
On the other hand, the smooth contribution $W_0$ remains arbitrary; once $W_0$ is fixed, it uniquely identify the state.

The subtraction of the Hadamard parametrix
defines the normal-ordered quantity $:G_k: := G_k - H_k$, smooth in the coincidence limit; the FRGE for $\Gamma_k$ thus becomes
\begin{multline} \label{eq:RG-Hadamard}
\partial_k \Gamma_k(\bar g; \phi) = \\ - \frac{1}{16\pi^2 \zeta^2_k} \int_\mathcal M \Tr{\partial_k \qk(x) \left [S_0 + V_0 \log{\frac{M^2}{\mu^2}} \right ] } \ . \\
\end{multline}
The logarithmic term $\log M^2$ is a smooth contribution coming from the arbitrary function $W$, and it is necessary to make the logarithm in Eq. \eqref{eq:G_k-Hadamard} dimensionless; $S_0 $ is the remaining smooth contribution in the coincidence limit.

Eq. \eqref{eq:RG-Hadamard} holds for a local regulator, and a Hadamard interacting propagator $G_k$. In Euclidean space, it is possible to derive a completely analogous equation, with the fundamental difference that the smooth contributions in the r.h.s. of Eq. \eqref{eq:RG-Hadamard} are uniquely fixed by Eq. \eqref{eq:def-G_k}. Moreover, in Euclidean space the coefficients $V_n$ can be equivalently computed by heat kernel techniques. However, in Lorentzian spacetimes the heat kernel is ill-defined, and discard state-dependent contributions \cite{Hack2012}.

\section{Einstein-Hilbert truncation}
The \textit{Einstein-Hilbert truncation} assumes an Ansatz for the effective average action in the form
\begin{multline}
\Gamma_k(\bar g; \phi, \sigma, \eta) = \Gamma_k^{EH}(\bar g, g) + \Gamma_k^{gh}(\bar g, h, c, \bar c) \\
+ \Gamma_k^{gf}(\bar g, h, b, c, \bar c) + \Sigma(\bar g; \phi, \sigma) + H(\bar g; \phi, \eta) \ .
\end{multline}
The Einstein-Hilbert contribution is
\[
\Gamma_k^{EH} = 2 \zeta_k^2 \int_\mathcal M \sqrt{-\det g} ( R(g) - 2 \Lambda_k) \ .
\]
In terms of the fluctuation field $h:= \langle \h \rangle = g - \bar g$, the ghost and gauge-fixing terms are
\begin{align*}
\Gamma_k^{gh} &= \zeta_k^2 \int_\mathcal M \sqrt{- \det \bar g} \bar c_a (\bar g^{ab} \square + \R^{ab}(\bar g)) c_b \ , \\
\Gamma_k^{gf} &= - \zeta_k^2 \int_\mathcal M \sqrt{- \det \bar g} b^a (\nabla^b h_{ab} - \frac{1}{2}\nabla_a \g^{bc}h_{bc} ) \ ,
\end{align*}
and $\Sigma$ and $H$ correspond to the classical contributions.

The equations for the interacting propagators are derived expanding  the effective average action up to second-order in a Taylor expansion,  $\Gamma_k(\bar g + h) = \Gamma_k(\bar g) + \mathcal O(h) + \Gamma_k^{\text{quad}}(h, \bar g)$. 

We can now specify the regulator terms $\qk, \ \qgh$. They are chosen to act as artificial masses for the fields, dressing the d'Alembertians as $\square \to \square - k^2$;
\begin{align}
\tensor{{\qk}}{^{ab}_{cd}} =  \zeta^2_k k^2 \tensor{K}{^{ab}_{cd}} \ , \quad \tensor{{\qgh}}{_{ab}} = \zeta^2_k k^2 \g_{ab} \ ,
\end{align}
where $K_{abcd} = 1/2 (\g_{ac} \g_{bd} + \g_{bc} \g_{ad}  - \g^{ab} \g_{cd})$.

The graviton propagator may be decomposed in the sum of a tensor $G_k^{T}$ and a scalar $G_k^S = \g^{ab} \g_{c'd'} \tensor{{G_k^{hh}}}{_{ab}^{cd}}$ contribution \cite{Allen1988}.
The equations of motion then read
\begin{align} \label{eq:G_k-EH-1}
 \zeta^2_k  [ &\g_{ac}\g_{bd} \left ( \square - k^2 + 2\Lambda_k  - \frac{1}{2} \R \right ) - \tensor{P}{_{ab}_{cd}} ] \tensor{{G_k^{T}}}{^{ab}^{c'd'}} \\ \nonumber
&= - \frac{1}{2}\left ( \tensor{\g}{_c^{c'}} \tensor{\g}{_d^{d'}} + \tensor{\g}{_d^{c'}} \tensor{\g}{_c^{d'}} - \g_{cd} \g^{c'd'} \right ) \delta(x,y) \\
\label{eq:G_k-EH-trace}
&- \frac{\zeta^2_k}{2} ( \square - k^2 + 2\Lambda_k ) G_k^S = - \delta(x,y) \ ,
 \\ \label{eq:G_k-EH-2}
& \zeta_k^2  \left [\g_{ab} (\square -k^2) +  \R_{ab} \right ] \tilde G_k^{ab'} =
- \tensor{\g}{_b^{b'}} \delta(x,y) \ .
\end{align}
The tensor $\tensor{P}{_{ab}^{cd}} := -2\tensor{\R}{_{(a}^{c}_{b)} ^{d}} - 2 \tensor{\g}{^{(c}_{(a}} \tensor{\R}{^{d)}_{b)}} + \g^{cd}\R_{ab} + \g_{ab}\R^{cd}$ is a potential term. In the last relation, all curvature quantities are constructed from the background metric $\bar g$; the d'Alembertian is $\square = \bar g(\nabla, \nabla)$. 

Each propagator has a corresponding Hadamard expansion:
\begin{align} \label{eq:G_k-Hadamard}
& G_k^S = - \frac{i}{ 4 \pi^2\zeta^2_k}  \left \{ H^S_k  + V^S_0 \log M^2_S + S^S_0 \right \} \\
& \tensor{{G^T_k}}{^{ab}^{c'd'}} = \frac{i}{ 8 \pi^2\zeta^2_k}  \left \{ \tensor{{H^T_k}}{^{ab}^{c'd'}}  + \tensor{{V_0^{T}}}{^{ab}^{c'd'}} \log M^2_{T} +  \tensor{{S_0^{T}}}{^{ab}^{c'd'}} \right \}  \\
& \tilde G_k^{ab'} = \frac{i}{ 8 \pi^2\zeta^2_k} \left \{ \tilde H_k^{ab'} + \tensor{{\tilde{ V_0}}}{^{ab'}} \log \tilde M^2 + \tensor{{\tilde{ S_0}}}{^{ab'}} \right \} \ .
\end{align}
The terms $V^{T}_0$, $V^S_0$, and $\tilde V_0$ arising from the equations \eqref{eq:G_k-EH-1}-\eqref{eq:G_k-EH-2} can be computed from Eq. \eqref{eq:V0} and are given by \cite{Allen1988, Belokogne2015}
\begin{align}\label{eq:V_0-explicit-1}
V_0^S &= \frac{1}{2}(k^2 - 2\Lambda_k) - \frac{1}{12}\R \\ \label{eq:V_0-explicit-2}
\tensor{{V^{T}}}{_0_{ab}^{cd}} &= - \frac{1}{12} \R \tensor{K}{_{ab}^{cd}} + \frac{1}{2}(\tensor{P}{_{ab}^{cd}} - \frac{1}{2} \g^{cd} \tensor{P}{_{abe}^e} )\\ \label{eq:V_0-explicit-3}
\tensor{\tilde V}{_0^{ab}} &= - \frac{1}{12} \g^{ab} \R + \frac{1}{2}(k^2 \g^{ab} - \R^{ab} ) \ .
\end{align}

\subsection{Universal terms and state dependence}
The RG equations \eqref{eq:RG-Hadamard} depends on the choice of a state. This is the main difficulty in applying the Lorentzian RG equations, in comparison with their Euclidean counterpart. In particular, Hadamard states for the graviton are not known in general spacetimes, but only in specific geometries \cite{Allen1986b, Allen1986c, Fewster2012, Benini2014, Gerard2022,Tsamis1992, Frob2016}. The construction of a Hadamard vacuum state for the graviton is well beyond the scope of this short note. Thus, here we take into account only universal contributions to the evolution equations, that are present in any Hadamard state and in all backgrounds. The evaluation of state-dependent contributions is possible only selecting a class of backgrounds, and it will be addressed in future works.

To solve the FRGE \eqref{eq:RG-Hadamard}, we need to evaluate $S_0 = \{ S_0^S , \ S^T_0,  \tilde S_0 \}$, and  $M^2_S, \ M^2_T  \ \tilde M^2$. First of all, the smooth functions $S_0$ vanish in the flat space limit \cite{Allen1986c, Hunt2012}. Moreover, any $k-$independent term can be removed by a re-definition of the effective average action, while terms proportional to the scale $k$ can be removed by an appropriate choice of the renormalization ambiguities \cite{HollandsWald2001a, HollandsWald2001b, HollandsWald2004}. Since the remaining contributions are completely state dependent, here we neglect $S_0$.

On the other hand, while the specific expressions for the functions $M^2_{T}, \ M^2_S, \ \tilde M^2$ are state-dependent, they must be present in any Hadamard state. They are functions of mass dimension 2, analytic in the physical parameters. The only dimension-2 term in the Hadamard expansion for the interacting propagator is $V_0$; we thus choose
\begin{equation}\label{eq:masses}
M^2_{S} = V_0^S \ , \quad M^2_{T} = \tensor{{V^{T}}}{_0_{ab}^{cd}} \tensor{I}{^{ab}_{cd}} \ , \quad \tilde M^2 = \tensor{\tilde V}{_0^{ab}} \g_{ab} \ ,
\end{equation}

These choices completely fix $W_0^S, \ W^T_0 \ , \tilde W_0$ and thus they fix a vacuum-like state through the Hadamard recursion relations. In the case of the scalar field, this choice coincides with the Minkowski vacuum state \cite{DDPR2022}.

The last term to be fixed is the arbitrary mass $\mu$. Contrary to the mass terms $M^2_{T}, \ M^2_S$, and $\tilde M^2$, depending on the choice of the state, this term is actually an arbitrary mass contribution coming from the choice of the Hadamard parametrix. Thus, we are free to choose a running Hadamard mass $\mu = k^2$, adjusting the UV regularisation to the renormalisation scale $k$. 

With these choices, the FRGE \eqref{eq:RG-Hadamard} is written in terms of state-independent, universal quantities.
%
%\begin{multline}
%k\partial_k \Gamma_k = 
%-\frac{k^2}{192 \pi ^2 \zeta^2_k} \bigg \{ (2 \zeta^2_k+\partial_k \zeta_k^2) \left(\left(6 k^2-12 \Lambda_k-\R\right) \log \left(\frac{1}{2}-\frac{\Lambda_k+\frac{\R}{12}}{k^2}\right)+4 \left(12 k^2-24 \Lambda_k+7 \R\right) \log \left(\frac{\frac{7 \R}{3}-8 \Lambda_k}{k^2}+4\right)+4 \left(12 k^2-5 \R\right) \log \left(2-\frac{5 \R}{6 k^2}\right)\right) \bigg \} \ .
%\end{multline}
Of course, state-dependent terms in specific backgrounds can significantly alter the FRGE.

\subsection{Phase diagram}
We can now compute the $\beta-$functions for the dimensionless constants $g_k$ and $\lambda_k$, related to the dimensionful running Newton's and cosmological constants by canonical rescalings:
\[
(32 \pi \zeta_k^2)^{-1} = G_k =  k^{-2} g_k \ , \quad \Lambda_k = k^2 \lambda_k \ .
\]
%The absolute value allows a flow to negative values in $G_k$.

Substituting the values for the $V_0$ coefficients, Eqs. \eqref{eq:V_0-explicit-1}-\eqref{eq:V_0-explicit-3}, the mass functions Eq. \eqref{eq:masses}, and setting the smooth contributions $S_0$ to $0$ in the RG flow \eqref{eq:RG-Hadamard}, we get a flow equation for the EAA written in terms of the Ricci scalar $\R$ and the coupling constants $\zeta^2_k$ and $\Lambda_k$.
% We can now expand the r.h.s of Eq. \eqref{eq:RG-Hadamard} up to first order in the background Ricci scalar $\R$. 
Notice that, thanks to the truncation, the r.h.s. of Eq. \eqref{eq:RG-Hadamard} depends on spacetime points only through the cut-off function $f$, and the trace is a simple trace over Lorentz and field indices. Thus, the functional derivatives on both sides of Eq. \eqref{eq:RG-Hadamard}, with respect to $\sqrt{-\text{det}g}$ and with respect to $\R$ at vanishing background fields, give the zeroth and first order in the Ricci scalar expansion, resulting in the evolution equations for $\zeta^2_k \Lambda_k$ and $\zeta^2_k$, respectively. The evolution equations are proportional to $f$, and we can take the adiabatic limit $f=1$ over the whole spacetime $\mathcal M$. Substituting the dimensionless coupling constants then give the $\beta-$functions for the dimensionless couplings $g_k$ and $\lambda_k$,
\begin{align} \label{eq:beta-functions-1}
k \partial_k g_k &= (\eta_{\text{N}} +2) g_k \\ \label{eq:beta-functions-2}
k \partial_k \lambda_k &=
-(2 - \eta_{\text{N}}) \lambda_k 
+ \frac{g_k}{4\pi} (2- \eta_{\text{N}}) \bigg \{  4 \log 4 \\ \nonumber
&+ (1 - 2\lambda_k) \left [ 8 \log[4(1-2\lambda_k)] + \log[\frac{1}{2}(1-2\lambda_k)] \right ]  \bigg \} \ ,
\end{align}
in terms of the anomalous dimension $\eta_{\text{N}} := G_k^{-1} k \partial_k G_k$:
\begin{equation}
\eta_{\text{N}}(g_k , \lambda_k ) = \frac{g_k}{6\pi} \frac{ 27 \log (1-2 \lambda_k) +7 + 37 \log 2}{1 + \frac{g_k}{12 \pi} \left ( 37 \log 2 + 27 \log(1-2\lambda_k) \right )} \ .
\end{equation}

The flow exhibits one non-trivial fixed point for $g_* = 1.15 , \ \lambda_* = 0.42$, realising the analogue of the Reuter fixed point in Lorentzian spacetimes. 
%The product $g_* \lambda_* = 0.48$, known to be more stable under changes in the renormalisation scheme, can be compared to the values found in the Euclidean case \cite{Reuter2001, ReuterSaueressig2019}. 
The critical coefficients for the Lorentzian fixed point are a pair of complex conjugate values, $\theta_{1,2} = 5.11 \pm 11.59 i$; therefore $\lambda_k$ and $g_k$ are two relevant directions, agreeing again with Euclidean results. These values can be compared to those obtained in the ADM formalism in Ref. \cite{Manrique2011}, that are $(g_*^{ADM}, \lambda_*^{ADM} ) = (0.21, 0.3)$ and $\theta_{1,2}^{ADM} = 0.94 \pm 3.1 i$. The Euclidean values are \cite{Lauscher2001} $(g_*^{E}, \lambda_*^{E} ) = (0.34, 0.3)$ and $\theta_{1,2}^E = 1.55 \pm 3.83 i$.  While the numerical values are roughly of the same order of magnitude, their difference is expected from the different spacetime signatures (Lorentzian v. Euclidean), choice of regulators (local v. non-local), and computational technique (Hadamard expansion scheme v. heat kernel techniques). 

In the ADM formalism of Ref. \cite{Manrique2011}, the interacting propagator is computed from a resummation of Matsubara frequencies in the compact time direction. The difference between the ADM-based formalism and the covariant formalism presented here should lie in different reference states $\omega$. In fact, the smooth contributions $W$ selecting a Hadamard state are related to the choice of positive frequencies along a selected time direction. The resummation of Matsubara frequencies suggests that the computation in Ref. \cite{Manrique2011} is performed with respect to a thermal KMS state at finite inverse temperature $\beta = k$. The computation presented here instead captures universal contributions to the RG flow, that are present in any state and in any background. The similarity between the two phase diagrams suggests that the choice of a thermal state can alter the precise values of the coupling constants at the fixed point and the critical exponents, but it leaves unaltered the existence and qualitative features of the fixed point.

The detailed connection between the two formalisms will be performed in future works, to highlight state and background dependence of the RG flow in Lorentzian spacetimes.
However, the qualitative picture of a non-Gaussian fixed point in the positive quadrant with critical exponents arises in all cases. The fixed point $(g_*, \ \lambda_*)$ thus provides a realisation of the AS scenario in Lorentzian spacetimes.

\begin{figure}[h]
\centering
\label{fig:fixed-point}
\includegraphics[width = \linewidth]{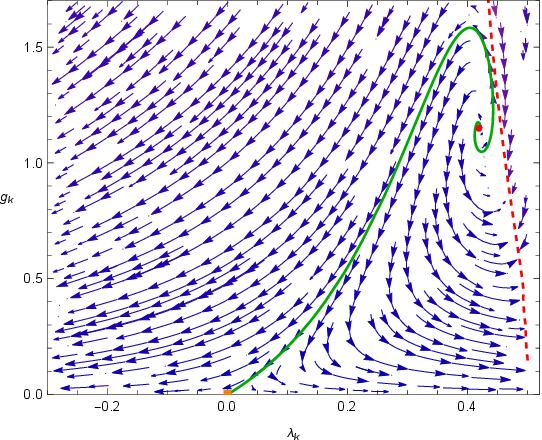}
\caption{Phase diagram obtained by numerical integration of the $\beta-$functions \eqref{eq:beta-functions-1}-\eqref{eq:beta-functions-2}.
%, with the distinguished \textit{Great Wave off Kawanaga} shape \cite{Hokusai}. 
The solid line is the separatrix, connecting the non-Gaussian fixed point (circle) to the Gaussian one (square); the dashed line denotes the locus where $\eta_{\text{N}}$ diverges.}
\end{figure}

\section{Conclusions}
The novel RG framework allows for the investigation of Lorentzian flows in a non-perturbative regime for gravity. In this note, we have seen that the contribution of universal, background independent terms in the flow of the Einstein-Hilbert truncation supports the evidence that gravity is non-perturbatively renormalisable also in the Lorentzian case.

To preserve background independence, we have restricted our attention to contributions to the flow coming only from universal terms. The important question now is if the non-trivial fixed point persists when state-dependent terms are taken into account. The investigation of state-dependent terms, however, requires to select a background. The Lorentzian FRGE \eqref{eq:RG-Hadamard} then allows for a systematic investigation of these state-dependent contributions in specific background geometries.

The RG flow state dependence can also be put in contact with the different runnings of the coupling constants in the Effective Field Theory approach to quantum gravity \cite{Donoghue2020}. In fact, the Newton's constant and the cosmological constant have different scalings in different scattering processes. Since the Lorentzian RG flow is state dependent, it is possible to study the flow of couplings in different states non-perturbatively.

The new formalism is tailored to Lorentzian spacetimes. The Hadamard expansion allows for quick generalisations to more advanced truncations. Universal terms in particular can be easily computed from the $V_0 , \ \tilde V_0$ terms in the Hadamard expansion in terms of the EAA thanks to Eq. \eqref{eq:V0}. The use of a local regulator and the Hadamard expansion of the interacting propagator allow for a relatively simple computation scheme for the contributions to the FRGE, preserving general covariance.
% The computation presented here can be compare to Ref. \cite{Manrique2011}, the only previous evaluation of the RG flow and of the AS scenario in Lorentzian spacetimes. There, it is necessary to assume a foliation of the background with a compact time direction, use heat kernel techniques to compute operator traces, and resum Matsubara frequencies, with a particular choice of non-local regulator.

The main novel result is that, in all backgrounds and for all Hadamard states, universal contributions are sufficient to identify a non-trivial fixed point in the RG flow, thus providing a universal mechanism for Asymptotic Safety in quantum gravity, at least in the Einstein-Hilbert truncation. The result is of particular relevance in a Lorentzian context, where there is an infinite family of interacting propagators for any given effective average action, indexed by a smooth function. Whether the choice of specific backgrounds and states can significantly alter this mechanism will be addressed in future works. 

Finally, while the EAA is a gauge-dependent quantity, gauge-invariant relational observables have been already studied in the context of locally covariant QG \cite{Brunetti2016b, Frob2017, Frob2018b, Frob2021} and in Euclidean fRG flows \cite{Baldazzi2021}. In future works, we plan to investigate the RG flow of gauge-invariant observables in Lorentzian quantum gravity.

\begin{acknowledgments}
I would like to thank Renata Ferrero, Paolo Meda, and Nicola Pinamonti for useful discussions. I am supported by a PhD scholarship of the University of Genoa and by the GNFM-INdAM Progetto Giovani \textit{Non-linear sigma models and the Lorentzian Wetterich equation}, CUP\textunderscore E53C22001930001.
\end{acknowledgments}

%\bibliographystyle{apsrev4-2}
%\bibliography{/Users/edoardodangelo/Documents/Fisica/Projects/bibliography-master.bib}

\end{document}